\documentclass[
    ,final            
  ]
  {aipproc}

\layoutstyle{6x9}

\def\propta{\hbox{${\propto}\!\!\!\!^{\sim}$}}
\def\la{\mathrel{\hbox{\rlap{\hbox{\lower4pt\hbox{$\sim$}}}\hbox{$<$}}}}

\begin{document}

\title{General Properties of Quiescent Novae}

\author{Brian Warner}{
  address={Department of Astronomy, University of Cape Town, Rondebosch 7700, 
South Africa}
}

\begin{abstract}
      
The observed properties of novae before and after eruption are 
discussed. The distribution of orbital periods of novae shows a 
concentration near 3.2 h, which resembles that of magnetic cataclysmic 
variables, and there is some evidence that many of the novae themselves are 
magnetic near that orbital period. Desynchronisation of polars by nova 
eruptions can lead to an estimate  ($\sim 2 \times 10^3$ y) for the time between 
eruptions for the strongly magnetic systems; this is much shorter than that 
found from other methods.
The similarity of pre- and post-nova luminosities, at high rates of 
mass transfer, is ascribed to irradiation of the secondary producing a 
self-sustained high $\dot{M}$ state. This slows cooling of the white dwarf after 
eruption, delays the onset of full scale dwarf nova outbursts in most 
systems, and delays any descent into a hibernation state of low rate of mass 
transfer.
\end{abstract}

\maketitle

\section{Introduction}

This conference is mostly about the high luminosity state, and the 
transitions into and out of it, but full understanding 
of the nova process must include the nature of the low luminosity white 
dwarf and its accretion environment, in which it spends most of its life.

\section{CYCLIC EVOLUTION AND HIBERNATION}

    As is well known, the mechanisms of orbital angular momentum loss 
generally invoked to drive CV evolution are magnetic braking (MB) for 
orbital periods longer than about 2 h and gravitational radiation (GR) for 
shorter periods (Note that it has long been pointed out that GR alone is not 
sufficient to account for the observed luminosities of many of the short 
period systems (Warner 1987), though these may not be representative of the 
long-term mean brightness). Basic models of MB (e.g. Verbunt \& Zwaan 1981) 
are single valued, giving a unique relationship between mass transfer rate 
($\dot{M}$) and orbital period $P_{orb}$. The large range (factors of 1000 or more) of 
$\dot{M}$ that is observed at most values of $P_{orb}$ (Patterson 1984; Warner 1987, 
1995a) shows that at least one other parameter is determining the 
instantaneous $\dot{M}$. Whether standard MB and GR set the long-term average, or 
whether additional mechanisms are required, is not yet certain. Large 
temporary excursions of $\dot{M}$ above the average set by MB can be generated by 
the effect of irradiation of the secondary by the primary and inner disc, 
which increases the scale height of the atmosphere of the secondary. If the 
response of the entire secondary to irradiative heating at the surface is 
calculated, cyclical evolution is found which alternates between a high $\dot{M}$ 
state in which the secondary expands and a low $\dot{M}$ state in which it 
contracts (King et al. 1995). Time scales for the transitions are $\sim 10^4 - 10^5$ 
y, and durations in the high states are $\sim 10^7$ y. Such cycles do not develop 
for secondary masses below $\sim 0.65$ M$_{\odot}$ because of the large thermal inertia 
in the convective envelope (King et al. 1996).

In addition, the observed high $\dot{M}$ after nova eruption, and the 
subsequent steady reduction, which is in quantitative agreement (e.g., for 
V1500 Cyg: Somers \& Naylor 1999) with the prediction (Prialnik 1986) of the 
effects of irradiation of the disc and the secondary by post-nova cooling of 
the white dwarf (Schreiber \& G\"ansicke 2001), show that variation about the 
mean certainly occurs through the intervention of novae. The additional mass 
transfer caused by heating of the secondary during (and for a century or 
more after) a nova eruption is, in theory, followed by a phase of lowered 
accretion rate (Kovetz, Prialnik \& Shara 1988) which is needed to maintain 
the secular average (but allowing for the fact that the eruption itself 
alters the orbital separation and therefore changes the radius of the Roche 
lobe of the secondary).
    
The question remains as to whether this is sufficient to send CVs into 
deep and sustained hibernation. Some evidence that this does not happen for 
at least 200 y has been given by Somers, Mukai \& Naylor (1996) and Somers, 
Ringwald \& Naylor (1997) for WY Sge (Nova Sagittae 1783), which is the 
oldest definitely recovered nova and still shows enhanced $\dot{M}$. But T Sco, 
the nova of 1860 that occurred in the globular cluster M 80, has been 
recovered and appears about a factor of ten fainter than normal nova 
remnants and does not appear (from absence of orbital modulation) to be a 
high inclination system (Shara \& Drissen 1995). In addition, not all novae 
that occurred early in the twentieth century have been recovered, and until 
they are it cannot be claimed that there is no evidence for hibernation 
within the first century or so of nova eruption. The most important fact 
continues to be that no stellar remnants of the bright novae noted in 
Oriental records of one to two millennia ago are observable today (Shara 
1989) -- instead of being the nova-like systems at magnitudes 12 -- 15 that 
modern bright naked eye novae become, they have faded beyond easy 
identification. This is the most persuasive case for eventual extended very 
low $\dot{M}$ states for post novae. 
     
Another case, and certainly easier to observe (at $V \sim 12$), is AE Aqr, 
with its magnetic primary rotating at $P_{rot}$ = 33 s, which must have been 
established as an equilibrium rotation period at high $\dot{M}$, but which has 
currently a low $\dot{M}$ suggestive of quite deep hibernation. Systems like AE 
Aqr would be quite difficult to discover at great distances (AE Aqr is one 
of the closest of CVs, at a distance of 86 pc), so they may be more common 
than realised.
    
It should also be kept in mind that at the shortest orbital periods 
there are dwarf novae like WZ Sge for which the estimated values of $\dot{M}$ are 
an order of magnitude or more below that set by GR, and so these systems are 
in at least partial hibernation. WZ Sge's spin period of $\sim$ 28 s (Warner \& Woudt 2002) also
indicates a high $\dot{M}$ in the past.

\section{PRE- AND POST-ERUPTION BEHAVIOUR}

    The survey of pre-eruptive behaviour of novae, made by Robinson (1975), 
disclosed only one (V446 Her) that could be thought to have had dwarf nova 
(DN)-like outbursts, with a range of nearly 4 mag. The variations were 
largely irregular, but the `flares' or `outbursts' had such slow rises to 
maxima ($\sim 10$ d) that they are incompatible with a CV having $P_{orb}$ = 4.97 h and 
(as concluded by Robinson on other grounds) are probably not normal DN 
outbursts. A similar remark may be made about the pre-outburst 1.6 mag 
variations in V3890 Sgr (Nova Sgr 1962: Dinerstein 1973). There is therefore 
no authenticated case of a nova eruption having taken place in a normal DN.
    
Robinson also found that pre- and post-eruptive magnitudes are very 
similar (an apparent exception, BT Mon, was later found to conform: Schaefer 
(1983)). This in general continues to be true, with definite exceptions of 
the three very fast novae GQ Mus (N 1983), CP Pup (N 1942) and V1500 Cyg (N 
1975), all of which rose from exceptionally faint magnitudes (to which they 
have not returned), though V1500 Cyg was in a brighter state for a week 
before eruption. Their pre-eruptive luminosities were so low that they were 
certainly in states of low $\dot{M}$ at those times.
    
The identity of pre- and post-eruptive luminosities is commonly used as 
evidence that a nova eruption does not seriously change the state of a CV. 
Yet the mass and angular momentum ejected in a nova eruption will certainly 
perturb the long-term orbital evolution -- the question is: How long will it 
take for the effects to show themselves? This is relevant to the evidence or 
otherwise for hibernation, discussed above. Here I want to examine the 
implications of pre- and post-novae being, both spectroscopically and in 
luminosity, indistinguishable from nova-like variables (e.g. Chapter 4 of 
Warner 1995a).
    
The interrelationships between nova-likes, VY Scl stars and DN, as 
functions of $P_{orb}$, can be summarised as follows:

$\bullet$ The VY Scl stars (which are nova-likes showing randomly distributed 
states of 

\hspace{0.3cm} low $\dot{M}$), all lie roughly in the $P_{orb}$ range 3.0 -- 4.0 h.

$\bullet$ There are very few DN in the $P_{orb}$ range 3.0 -- 4.0 h.

$\bullet$ The values of $\dot{M}$ that appear among the nova-likes in the region of 
$P_{orb}$ = 3 -- 4 h 

\hspace{0.3cm} are more than an order of magnitude greater than that 
predicted by the theory 

\hspace{0.3cm} of magnetic braking (Warner 1987).

These strong correlations have been quantitatively explained by Wu, 
Wickramasinghe \& Warner (1995a,b, hereafter WWW; see also Warner 1995a) in 
the following way: For $P_{orb}$ < 4 h the separation between the secondary and 
primary is small enough to produce significant irradiative heating of the 
secondary by the hot central regions of the disc and the primary which 
results in greatly enhanced $\dot{M}$. The surface temperature $T_{eff}$   of the 
primary is largely governed by $\dot{M}$ -- in particular, the large values of 
$T_{eff}$ found in nova-likes (up to 50\,000 K, Sion 1999) are what are expected 
for accretion at $\dot{M} \sim 10^{-8}$ M$_{\odot}$ y$^{-1}$ (Section 9.4.4 of Warner 1995a). WWW 
found that, as the thickness of the accretion disc increases with $\dot{M}$ and 
shields the secondary, this negative feedback automatically leads to an 
upper limit of $\dot{M}$ $\sim$ few $\times 10^{-8}$ M$_{\odot}$ y$^{-1}$, and that the non-linearity of the 
situation leads to short lived high/low $\dot{M}$ states as observed in the VY 
Scl stars (in which the important time scales are the response time of the 
outer envelope of the secondary, and the cooling time of the outer parts of 
the primary) and/or to long-lived ($10^4 - 10^6$ y) high and low $\dot{M}$ 
excursions. The upper limit on $\dot{M}$ corresponds to M$_V \sim 3.3$ in the 3.0 < 
$P_{orb}$ < 4.0 h region, only slightly brighter than what is actually observed 
(Warner 1987).
    
To summarise: the predominance of nova-likes, and in particular the VY 
Scl stars, in the $P_{orb}$ = 3 -- 4 h range is probably due to 
irradiation-enhanced $\dot{M}$; the near absence of DN is due to the relative 
rapidity of passage through states of intermediate $\dot{M}$. Below $P_{orb} \sim 3$ h 
the mechanism causing the orbital period gap dominates.
     
The relevance of this theory to novae is clear: almost all novae are 
observed to erupt from CVs of the nova-like subtype, where $\dot{M}$ is high 
(naturally high for $P_{orb}$ > 4 h, enhanced by irradiation to a high value for 
$P_{orb}$ < 4 h) and where a primary with $T_{eff}$ up to $\sim$ 50\,000 K resides. After 
eruption, the primary has an even higher temperature, and $\dot{M}$ is thereby 
even more enhanced (Prialnik 1986), but both decrease as the primary cools. 
However, for $P_{orb} \la  4$ h, the primary is prevented from cooling 
below $T_{eff} \sim$ 50\,000 K because the irradiation-enhanced high $\dot{M}$ equilibrium 
is re-established after eruption. It is this effect that results  (at least 
for the shorter $P_{orb}$ systems) in equality of luminosity before and after 
eruption. The equilibrium value of $T_{eff}$ is sensitive to the mass of the 
primary (higher masses lead to more gravitational potential energy release 
but a smaller area to radiate it away); using equations 2.83b and 9.55 of 
Warner (1995a) we find $T_{eff}$ \propta $M^3$ for $M$ > 1.0 M$_{\odot}$, where a large primary 
mass is adopted because there is higher probability for nova eruptions to be 
observed in high mass systems. As the selection for higher masses does not 
apply to nova-likes we should find that $T_{eff}$, $\dot{M}$ (and hence M$_V$) in 
nova-likes is on average smaller (fainter) than in old post-novae. 
Comparison of Figures 4.16 and 4.20 of Warner (1995a) shows this 
is in fact the case.
    
For non-magnetic CVs with $P_{orb}$ < 4 h, therefore, the primary's cooling 
curve as computed by Prialnik (1986) applies only until the 
irradiation-enhanced pre-nova equilibrium is regained. How 
long the latter phase will last is not yet known. Clearly, the effects of 
mass and angular momentum lost during eruption eventually must take their 
toll -- in order to maintain the long-term average angular momentum drain 
from the orbit. For $P_{orb}$ < 4 h the irradiative equilibrium phase acts to 
delay the onset of the necessary low $\dot{M}$ state -- an example of deferred 
compensation.
    
These remarks are also relevant to the nature of the pre-eruption 
outbursts seen in V446 Her and V3890 Sgr, as mentioned above. The 
pre-eruption range of V446 Her was m$_{pg} \sim 14.9 - 18.4$ (Robinson 1975) -- and 
is larger than the range seen in its post-eruption variations, which are 
clearly standard DN outbursts (Honeycutt, Robertson \& Turner 1995, Honeycutt 
et al. 1998). This and the slow rises appear more like VY Scl behaviour, but 
the large $P_{orb}$ argues against this and it is likely that what is seen is a 
combination of normal DN outbursts (greatly under-sampled by the 
photographic archive plates) and some variations in $\dot{M}$, implying that $\dot{M}$ 
was low enough before (as after) eruption for the accretion disc to be 
thermally unstable. We note that at $P_{orb} \sim 5$ h the irradiation in V446 Her 
will not be sufficient to hold the system in a high $\dot{M}$ nova-like state if 
its natural $\dot{M}$ is below the critical value.
   
This last effect is seen as responsible for the non-appearance of DN 
outbursts in the majority of post-novae. The only systems for which 
authenticated standard DN outbursts have been seen are V446 Her ($P_{orb}$ = 4.97 
h), GK Per ($P_{orb}$ = 47.9 h) and V1017 Sgr ($P_{orb}$ = 137 h), for all of which 
irradiation of the secondary is not important because of the large 
separations implied by the long orbital periods.
   
On the other hand, `stunted' DN outbursts in nova-likes and old 
post-novae are commonly observed (Honeycutt, Robertson \& Turner 1998). In 
these the time scales of typical DN outbursts are seen, but the amplitudes 
are only $\sim 0.6$ mag. These can be understood as arising from outbursts in 
only the outer parts of the accretion discs, with the inner parts kept 
permanently in a high temperature state through irradiation by the hot 
primary (Warner 1995b; Schreiber, G\"ansicke \& Cannizzo 2000).
    
It should be mentioned that, in addition to the variations in brightness 
on DN time scales (weeks or months), there are low amplitude (typically 0.1 
-- 0.2 mag) variations on time scales of years believed to be caused by $\dot{M}$ 
variations resulting from magnetic cycling within the secondary (see Table 
9.3 of Warner 1995a).
    
To return to pre-eruption light curves, Robinson (1975) found that 5 out 
of 11 well observed systems showed slow increases of brightness of 0.25 -- 
1.5 mag during 1 -- 15 y before eruption. With the possible exception of the 
largest value (which is for V533 Her) the M$_V$ increases are modest and may be 
the result of increases in $\dot{M}$ from the secondaries as seen in the decadal 
cycles mentioned above. (Note, however, that an increase of only 0.33 mag in 
V for a high $\dot{M}$ disc implies an increase in $\dot{M}$ by a factor $\sim 2$ (Smak 
1989) -- this may not apply to V533 Her, which is an intermediate polar (see 
below) and has a truncated disc). Such an increase in $\dot{M}$, with its 
concomitant increase of compressional heating in the surface layers of the 
primary, could well trigger a nova eruption. The observations imply that 
about half of nova-likes that have accreted almost a critical mass are 
triggered during a high part of a decadal cycle. Of course, this is what 
would be expected randomly anyway, so it is not evidence for such a 
triggering mechanism!
     
Finally, high $\dot{M}$ discs in CVs with $P_{orb} \la 4$ h commonly 
show superhumps arising from precessing elliptical accretion discs (e.g. 
Patterson 1999), and post-novae are no exception. V603 Aql, V1974 Cyg, CP 
Pup and probably V4633 Sgr, V2214 Oph and GQ Mus are examples. Superhumps 
are a diagnostic for high $\dot{M}$ discs, and as such can be used to identify 
high luminosity discs at short $P_{orb}$ -- where almost all CVs have very low 
$\dot{M}$ (e.g. Fig. 9.8 of Warner 1995a). The only one so far found (that is not 
a known recent nova) is BK Lyn ($P_{orb}$ = 0.075 d; Skillman \& Patterson 1993), 
which may possibly be the remnant of Nova Lyn 101 AD (Hertzog 1986). Such 
systems have very low amplitude photometric modulations and are difficult to 
find. The only known reason for high $\dot{M}$ at such short $P_{orb}$ is connected 
with eruption, and such systems therefore are strongly indicative of 
prehistoric novae. If the identification of BK Lyn with the nova of 101 AD 
could be proven, it would show that at very short $P_{orb}$ a high $\dot{M}$ might be 
(self-) sustained for at least a millennium.

\section{ORBITAL PERIODS OF NOVAE}

In their discussion five years ago of the orbital period distribution of 
novae, and their progenitor population, Diaz and Bruch (1997) listed 30 
objects with known $P_{orb}$ < 24 h (six of which are considered by Downes et al. 
to be unreliable period determinations). The number of classical novae 
(omitting recurrent novae, all of which seem to be different from the 
`non-recurrent' ones) with known periods < 24 h is now the 50 listed in 
Table I (which also omits some uncertain determinations - and note that a 
few may be superhump periods, which are a few percent different from $P_{orb}$), 
and demonstrates considerable observational progress in the past 5 years.

\begin{table}
\begin{tabular}{lcclclccl}
\hline
\tablehead{1}{l}{b}{Star} &
\tablehead{1}{c}{b}{Date} &
\tablehead{1}{c}{b}{Magn. range} &
\tablehead{1}{l}{b}{$P_{orb}$} &
\tablehead{1}{l}{b}{\ \ } &
\tablehead{1}{l}{b}{Star} &
\tablehead{1}{c}{b}{Date} &
\tablehead{1}{c}{b}{Magn. range} &
\tablehead{1}{l}{b}{$P_{orb}$} \\
\hline
RW UMi     &1956&   6 -- 18.5 &   1.418 & & WY Sge       &1783&   5.4 -- 20.7 &   3.687 \\
GQ Mus     &1983& 7.2 -- 18.3 &	  1.425 & & OY Ara       &1910&   6.0 -- 17.5 &   3.731 \\
CP Pup     &1942& 0.5 -- 15.2 &   1.474 & & V1493 Aql    &1999&	 10.4 -- >21  &   3.74  \\
V1974 Cyg  &1992& 4.2 -- 16.1 &   1.950 & & V4077 Sgr    &1982&   8.0 --  22  &   3.84  \\
RS Car     &1895& 7.0 -- 18.5 &   1.980 & & DO Aql       &1925&   8.7 -- 16.5 &   4.026 \\
DD Cir     &1999& 7.7 -- 20.2 &   2.340\tablenote{Woudt \& Warner, unpublished} & & V849 Oph     &1919&	  7.3 -- 17   &   4.146 \\
V Per      &1887& 9.2 -- 18.5 &   2.571 & & V697 Sco     &1941&  10.2 -- 19.7 &   4.53$^*$ \\
QU Vul     &1984& 5.6 -- 17.5 &   2.682 & & DQ Her       &1934&   1.3 -- 14.6 &   4.647 \\
V2214 Oph  &1988& 8.5 -- 20.5 &   2.804 & & CT Ser       &1948&   7.9 -- 16.6 &   4.68  \\
V630 Sgr   &1936& 1.6 -- 17.6 &   2.831 & & T Aur        &1891&   4.2 -- 15.2 &   4.906 \\
V351 Pup   &1991& 6.4 -- 19.0 &   2.837 & & V446 Her     &1960&   3.0 -- 17.8 &   4.97  \\
V4633 Sgr  &1998& 7.4 -- >20  &   3.014 & & V533 Her     &1963&   3.0 -- 15.0 &   5.04  \\
DN Gem     &1912& 3.5 -- 16.0 &   3.068 & & HZ Pup       &1963&   7.7 -- 17.0 &   5.11  \\
V1494 Aql  &1999& 4.0 -- >16  &   3.232 & & AP Cru       &1936&  10.7 -- 18.0 &   5.12  \\
V1668 Cyg  &1978& 6.7 -- 19.8 &   3.322 & & HR Del       &1967&   3.5 -- 12.3 &   5.140 \\
V603 Aql  &1918&--1.1 -- 11.8 &   3.324 & & V1425 Aql    &1995&   7.5 -- >19  &   5.419 \\
DY Pup     &1902& 7.0 -- 19.6 &   3.336 & & BY Cir       &1995&   7.2 -- 17.9 &   6.76$^*$ \\
V1500 Cyg  &1975& 2.2 -- 18.0 &   3.351 & & V838 Her     &1991&   5.4 -- 15.4 &   7.143 \\
V909 Sgr   &1941& 6.8 -- 20   &   3.36  & & BT Mon       &1939&   8.5 -- 16.1 &   8.012 \\
RR Cha     &1953& 7.1 -- 18.4 &   3.370 & & V368 Aql     &1936&   5.0 -- 15.4 &   8.285 \\
RR Pic     &1925& 1.0 -- 12.1 &   3.481 & & QZ Aur       &1964&   6.0 -- 17.5 &   8.580 \\
V500 Aql   &1943& 6.6 -- 17.8 &   3.485 & & CP Cru       &1996&   9.2 -- 19.6 &  11.3$^*$  \\
V382 Vel   &1999& 2.7 -- 16.6 &   3.508 & & DI Lac       &1910&   4.6 -- 15.0 &  13.050 \\
V533 Her   &1963& 3.0 -- 14.8 &   3.53  & & V841 Oph     &1848&   4.2 -- 13.5 &  14.50  \\
V992 Sco   &1992& 8.3 -- 17.2 &   3.683$^*$ & & V723 Cas     &1995&   7.1 -- >18  &  16.638 \\
\hline
\end{tabular}
\caption{Orbital Periods of Novae}
\label{tab1}
\end{table}
    
\begin{figure}
  \includegraphics[height=.3\textheight]{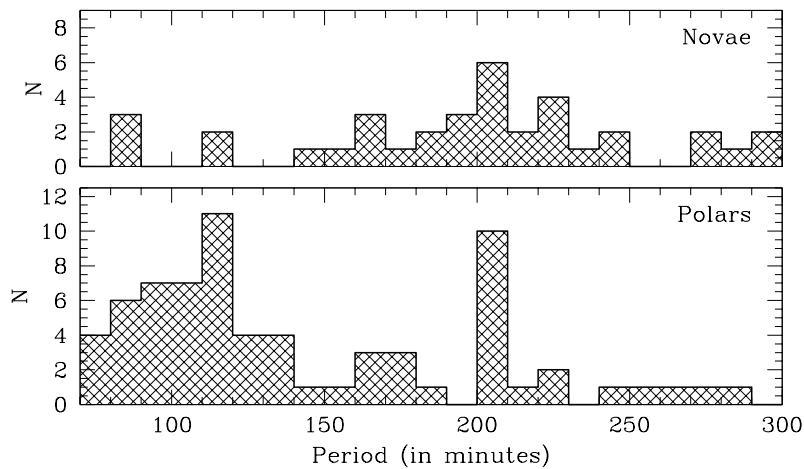}
  \caption{The frequency distribution of the orbital periods of Novae (upper panel) and Polars (lower panel).}
  \label{fig2}
\end{figure}

The frequency distribution of nova orbital periods is shown in Figure~\ref{fig2}.
    
To the list of novae that have been observed to erupt it should 
eventually be possible to add ones that are demonstrably novae that were 
overlooked in the relatively recent past. As a class, the desynchronised
polars, discussed below, provide four probable examples with periods 1.85, 
3.35, 3.35 and 3.37 h. The X-Ray source RX\,J1039.7-0507, with $P_{orb}$ = 1.574 
h, is probably another (see poster by Woudt \& Warner at this conference).
   
A FAQ (Frequently Asked Question) is whether the $P_{orb}$ distribution for 
classical novae shows a gap (or, at least, a greatly lowered space density) 
in the range 2 -- 3 h in the same way as dwarf novae (see, e.g., Warner 
1995). Using the range 2.11 < $P_{orb}$(h) < 3.20 for the empirically determined 
period gap (for CVs of all types: Diaz \& Bruch, 1997) there are eight novae 
in the `gap', which does not support the presence of a gap -- though a 
reduction of population relative to the number of novae immediately above 
3.2 h is undeniable -- but so is the rapid fall in numbers for $P_{orb}$ below 3 
h.
    
A FUQ (Frequently Unasked Question) is what we should learn from the 
answer to the above FAQ. A point to be considered is that all white dwarfs 
accreting at the rates commonly seen in CVs must eventually undergo nova 
explosions (we exclude here the highest accretion rates which lead to Ultra 
Soft X-Ray Sources steadily burning hydrogen near their surfaces). The 
population of novae is therefore drawn from all of those CV subtypes that 
have high enough accretion rates to produce novae. If there is a population 
of detached systems in the period gap, which is the conventional explanation 
for the `missing' CVs in the gap, they obviously do not contribute to the 
census of novae. If there is in fact a lower space density of novae at the 
position of the traditional period gap, it may simply be representing the 
gap we already see in some other subtypes.

There are few known short orbital period novae, and the reason for this 
is not immediately obvious. The space density of CVs in the range $P_{orb}$ -- 
$P_{orb}$ + d$P_{orb}$ should be inversely proportional to d$P_{orb}$/d$t$, and therefore 
proportional to <$\dot{M}$>$^{-1}$, and the frequency of nova eruptions should be 
proportional to  <$\dot{M}$> (where <> denotes the long-term secular mean), so 
the fraction of CVs that become novae should be roughly independent of 
<$\dot{M}$>. The observed large pile-up of CVs at short $P_{orb}$ (which is even 
larger in distributions predicted by population syntheses) -- is not 
represented by a similarly large number of novae. Furthermore, the novae are 
sampled from a larger volume of space than the other CVs, so the short $P_{orb}$ 
novae have an even lower relative space density than expected.
     
Even if there is no obvious period gap, Table~\ref{tab1} and Fig.~\ref{fig2} do have one 
distinctive feature: there is a concentration of novae in the range 2.8 -- 
4.1 h (44\% of all known orbits lie in this 1.3 h range). The remainder are 
spread widely: another 48\% cover the 6.3 h of ranges 1.4 -- 2.8 h and 4.1 -- 9 
h. In some respects this is a mirror image of what happens in the DN, where 
there is a large number with $P_{orb}$ < 2 h and very few in the range 3 -- 4 h. 
We can partly understand this through the fact that the average mass 
transfer rate is very much lower below the period gap than above, which 
ensures that most CVs below the gap lie below the critical transfer rate for 
stable accretion discs, and at the same time they accrete mass so slowly 
that the frequency of nova eruption {\sl per star} is low. But above the gap the 
anticorrelation of classical and dwarf novae populations must be more 
subtle.  We shall see below that there is some evidence that many of the 
novae in the central part of the 3 -- 4 h range have magnetic primaries; 
these would have to be subtracted from any comparison with the dwarf novae 
(which at most are weakly magnetic), but this still leaves the 3 -- 4 h range 
clearly favoured by novae.
     
A possible explanation has already been given above in terms of the 
effects of irradiation of the secondary (WWW), which begin to be very 
important for $P_{orb}$ < 4 h. If mean values of $\dot{M}$ over times of $10^3 - 10^4$ y 
are either very high or very low in the 3 -- 4 h range, with transitions 
between taking place relatively rapidly (hundreds of years), then dwarf 
novae (which would only exist in the transition region) will be rare. The 
only evidence for possible secular change in $\dot{M}$ due to such transition is 
the difference in mean outburst intervals in U Gem ($P_{orb}$ = 4.25 h), which 
are 96.5 d for 1855--1905 and 107.6 d for 1905--1955 (Warner 1987).

\section{MAGNETIC NOVAE}
     
At least a quarter of CVs have primaries with magnetic fields strong 
enough to affect the accretion flow. The strongest fields, in the polars, 
prevent the formation of accretion discs; the intermediate polars and DQ Her 
stars, which have progressively lower field strengths, have accretion discs 
truncated at their inner edge by the magnetosphere of the primary.
    
Eruptions on magnetic white dwarfs have occurred in RR Cha, GK Per, HZ Pup and V697 Sco,
which are intermediate polars, and in DQ Her and V533 Her, but Nova Cygni 
1975 (V1500 Cyg) remains unique as the only observed nova eruption that has 
been proven to have arisen on a strongly magnetic white dwarf, i.e. a polar. 
Photometry 12 years after the eruption showed it to have a light curve like 
that of a polar (Kaluzny \& Semeniuk 1987) and subsequent polarimetric 
observations revealed the characteristics of a polar (Stockman, Schmidt \& 
Lamb 1988) with a field $\sim$ 25 MG. An unexpected feature, however, was that 
the rotation period of the white dwarf is shorter by 1.8\% than the orbital 
period -- which is interpreted as being the result of coupling between the 
expanded atmosphere of the primary and secondary during eruption, with 
subsequent spin-up as the atmosphere collapsed back later. Observations have 
shown that the spin period $P_{spin}$ is increasing at a rate that implies 
resynchronisation in $\sim$ 185 y (Schmidt, Liebert \& Stockman 1995).
   
Nova Puppis 1991 (V351 Pup) has recently been found to have a light curve 
very similar to that observed for V1500 Cyg a decade after its eruption; 
although not yet detectably polarized, it may well turn out to be another 
example of eruption of a strongly magnetic system (Woudt \& Warner 2001).
   
Indirect evidence for eruptions that occurred in magnetic systems in the 
past, but went unrecorded as novae, is given by the occurrence of three 
other desynchronised polars: BY Cam (Mason et al. 1998), V1432 Aql 
(RX\,J1940.1-1025) (Geckeler \& Staubert 1997) and CD Ind (RX\,J2115-5840) 
(Ramsay et al. 2000). Their properties are listed in Table~\ref{tab2}. The fact that 
$P_{spin}$ > $P_{orb}$ in V1432 Aql may be explained as coupling of the magnetic 
field of the primary with the dense wind during eruption (this is in 
competition with the purely dynamical transfer of angular momentum mentioned 
above for V1500 Cyg). The measured resynchronisation time scales, listed in 
Table~\ref{tab2}, range over an order of magnitude: $\sim 100 - 1000$ y, not correlated 
with the amount of asynchronism (as measured by the beat period $P_{beat}$ 
between the orbital and white dwarf spin periods).
    
V1432 Aql provides another form of indirect evidence for historical nova 
eruption. Schmidt \& Stockman (2001) measure an effective temperature of 
35\,000 K for the primary, which is much greater than the typical 8000 -- 20\,000 
K found for other polars (other than V1500 Cyg, which has 90\,000 K from its 
recent eruption) and is out of equilibrium with the present rate of 
accretion heating. From the cooling calculations made by Prialnik (1986) 
this indicates a nova eruption in the past 75 -- 150 y.
     
AE Aqr, already mentioned above, has an unknown $T_{eff}$  because the 
dominant measurable UV flux comes from spots with $T \sim 26\,000$ K produced by 
heating by the accretion columns (Eracleous et al. 1994). If the general 
non-heated surface temperature of the primary could be measured more 
accurately than the currently estimated 10\,000 -- 16\,000 K this would provide 
a measure of how long ago the high $\dot{M}$ phase (characteristic of the large 
$P_{orb}$ of AE Aqr) was interrupted, probably by a nova eruption.

\begin{table}
\begin{tabular}{lrrrr}
\hline
\tablehead{1}{l}{b}{Star} &
\tablehead{1}{r}{b}{$P_{orb}$ (mins)} &
\tablehead{1}{r}{b}{$P_{spin}$ (mins)} &
\tablehead{1}{r}{b}{$P_{beat}$ (d)} &
\tablehead{1}{r}{b}{$T_{syn}$ (y)} \\
\hline
V1432 Aql     &       201.94       &         202.51         &        49.5     &        110\\
BY Cam        &        201.26      &          199.33        &         14.5    &       1107\\
V1500 Cyg     &       201.04       &         197.50         &          7.8    &        185\\
CD Ind        &           110.89   &             109.55     &              6.3& \\
\hline
\end{tabular}
\caption{Desynchronised Polars}
\label{tab2}
\end{table}

As has been pointed out before (Warner 1995a), the fraction $f$ of polars 
that are desynchronised may provide a means of estimating observationally 
the otherwise inaccessible average time $T_R$ between nova eruptions, at least 
for the magnetic systems. If <$T_{syn}$> is the average resynchronisation time 
then $T_R$ = <$T_{syn}$>/$f$. There are about 68 known polars, so with 4 observed to 
be desynchronised and a mean <$T_{syn}$> $\sim$ 300 y we have $T_R \sim 5000$ y. This crude 
figure can be refined in several ways but, as we see below, it produces a 
serious conflict with what is known about $T_R$ from other directions.
   
First, $T_{syn}$ depends on a number of system parameters, including the mass 
of the white dwarf, the mass ejected and how well angular momentum was 
exchanged with it, the time since the eruption, etc. A global average of 
these effects could be obtained by appropriate theoretical modelling, but 
this may not be justified at present because of the small number of systems 
included in the statistics.
   
Second, $f$ is undoubtedly underestimated because of insufficient 
observational coverage of many of the fainter polars. Perhaps not even all 
the polars with m$_v$ < 16.0 have been studied enough to detect 
asynchronism, but assuming that they have we note that V1432 Aql, BY Cam and 
CD Ind (we exclude V1500 Cyg as having been discovered in a non-standard 
way) constitute 3 out of about 20 systems with high state magnitudes 
brighter than 16. If this fraction applies also to the total population of 
polars, then there are another 7 de-synchronised systems among the fainter 
members (one of which is already known: V1500 Cyg, at m$_v$ = 18.0). 
This gives $f \sim 0.15$ and reduces $T_R$ to 2000 y.
   
Third, there are difficulties in detecting asynchronism for the older 
magnetic novae where synchronism has been nearly re-established. In essence, 
when the beat period $P_{beat}$ becomes $\sim$ months there is an observational bias 
against finding such systems. Superficially, it might be thought that, as 
the observed values of $P_{beat}$ lie in the range 6 -- 50 d, leaving the range 50 
-- $\infty$ d unexplored, there could be a large fraction of currently undetected 
desynchronised polars. However, the following reasoning suggests that the 
loss is not great.
    
The key aspect is that the synchronization torque is independent of the 
amount of de-synchronisation and is constant with time. The general form of 
the torque $N_{syn}$ is $N_{syn} \sim \mu_1 \mu_2 / a^3$, where $\mu$ is the magnetic moment 
(indigenous or induced in the case of the secondary) and $a$ is the separation 
of stellar components (e.g. Hameury, King \& Lasota 1987). The 
synchronisation time is $T_{syn}$ = ($P_{orb}$ - $P_{spin}$)/$\dot{P}$, where $\dot{P}$ = 
d$P_{spin}$/d$t$  
is essentially constant with time because of the constant torque. At any 
time after the nova eruption we have the relationship

\begin{equation}
P_{beat} =   {{P_{orb}^2}\over{| \dot{P} | T_{syn}}} ,
\end{equation}

where $T_{syn}$ is the time remaining until synchronisation. Suppose that the 
initial de-synchronisation produces a beat period $P_{beat}$(init) (typically a 
few days) and that current observational techniques make it difficult to 
detect asynchronism for beat periods longer than some limit $P_{beat}$(limit) 
(typically a few weeks). Then the fraction $F$ of asynchronous systems that is 
detectable is

\begin{equation}
 F  =  {T_{syn}\,{\rm (init)} - T_{syn}\,{\rm (limit)}\over{T_{syn}\,{\rm (init)}}} =   1 - {P_{beat}\,{\rm (init)}\over{P_{beat}\,{\rm (limit)}}}.
\end{equation}

The result is that with $P_{beat}$(init) $\sim$ week and $P_{beat}$(limit) $\sim$ months, 
very few of the desynchronised systems are overlooked, so the difficulty of 
detecting differences between $P_{orb}$ and $P_{spin}$ when they are very small has 
little effect on the estimate of $f$.
     
The deduced value of $T_R \sim 2000$ y is distressingly incompatible with the 
estimated mass $\sim 2 \times 10^{-4}$ M$_{\odot}$ of ejecta in V1500 Cyg (Hjellming 1990) and 
the average mass transfer rate $\sim 10^{-9}$ M$_{\odot}$ y$^{-1}$ estimated for longer period 
polars (Beuermann \& Burwitz 1995), which give $T_R \sim 2 \times 10^5$ y. It could be 
that nova eruption is not the only mechanism that is capable of 
desynchronising polars, or that the ejecta masses of magnetic novae are 
grossly overestimated.  Weakening of the synchronising torque with time does 
not help because it puts a larger fraction into systems with large $P_{beat}$ 
that would currently be overlooked, necessitating an even larger correction 
to obtain the true number of desynchronised polars.
    
Another FAQ is whether polars show any period gap. The latter subject is 
one set about with controversy (Beuermann \& Burwitz 1995; Wickramasinghe \& 
Ferrario 2000). The list of polars shows that no absolutely empty gap is 
present. Beuermann \& Burwitz note that comparing the number of systems in 
the period range of the gap with that just outside shows that gap polars are 
relatively twice as frequent as non-magnetic gap systems. They say that 
nevertheless the orbital distributions of magnetic and non-magnetic systems 
are not statistically different. Another point of view could be that the 
`period gap' would probably not be noticed in polars 
if it were not so prominent in the non-magnetic CVs -- and that it is 
probably equally true that the polar distribution is not statistically 
different from one that has no gap at all. However, here it is necessary to 
distinguish between two kinds of polars -- several of the polars in the gap 
are strong cyclotron line emitters and are interpreted as having extremely 
low accretion rates  $\sim 10^{-13}$ M$_{\odot}$ y$^{-1}$ (Reimers \& Hagen 2000). Such rates are 
< $10^{-3}$ of what is normally seen in polars and constitute systems in which 
accretion has nearly shut down, as predicted by some explanations of the 
period gap. For present purposes these systems should therefore be removed 
from the census of polars in the gap. These particular gap-filling polars 
are not going to contribute to the census of novae -- they will take at least 
$10^8$ y to accumulate sufficient mass to trigger a nova eruption.
    
Novae are drawn from all those CV subtypes that have sufficiently high 
long-term average accretion rates. In the case of the magnetic systems we 
might look for similarities in the distributions of polars and novae. The 
frequency distribution of polars has spikes near 114 min (1.90 h) and 202 
min (3.37 h) (Wickramasinghe \& Ferrario 2001); it happens that all four of 
the desynchronised polars listed in Table~\ref{tab2} are members of these two groups.
     
Two novae (V1974 Cyg and RS Car) out of the five that have $P_{orb}$ < 2.0 h 
coincide with the 114 min peak. RS Car is strongly modulated in brightness 
but has not been observed well enough to detect any periods other than $P_{orb}$ 
(or a superhump period), but V1974 Cyg could be a desynchronised polar -- its 
multiple periodicities are interpreted as such by Semeniuk et al. (1995), but 
an explanation in terms of superhumps is preferred by Skillman et al. (1997) 
and Retter, Leibowitz \& Ofek (1997). On the other hand, Shore et al. (1997) 
find that the flux distribution in V1974 Cyg can be accounted for entirely 
by a hot white dwarf, leaving no evidence for the presence of the disc 
required to generate superhumps, and thus indirectly supporting the 
desynchronised polar model.
    
This may be pure coincidence, but it is interesting that in addition to 
the 114 min peak, the concentration of novae around 3.3 h is centred roughly 
on 202 min and includes the magnetic nova type specimen V1500 Cyg. This 
suggests looking for possible evidence of magnetism in other novae within 
the 3.3 h cluster -- the results are shown in Table~\ref{tab3}.

\begin{table}
\begin{tabular}{ll}
\hline
\tablehead{1}{l}{b}{Star} &
\tablehead{1}{l}{b}{Remarks} \\
\hline
V351 Pup	&Possibly magnetic like V1500 Cyg (Woudt \& Warner 2001).\\
V4633 Sgr	&Possible asynchronous polar (but also could simply be a\\
& superhump modulation) (Lipkin et al. 2001).\\
DN Gem          &No evidence for magnetism.\\
V1494 Aql &	41.7min X-Ray period attributed to pulsations (Starrfield \&
Drake 2001). \\
 & No direct evidence for magnetism.\\
V1668 Cyg & Well observed but no magnetic signature at quiescence
\tablenote{However, the apparently well-defined photometric modulation at a period of 
10.54 h, seen during early decline (Campolonghi et al. 1980), could be a spin 
beat with the orbital period. A cycle of this modulation was detected by Di 
Paolantonio et al. (1981) but not by some other observers (Piccioni et al. 
1984; Kaluzny 1990).}
.\\
V603 Aql  &	Possible intermediate polar, $P_{spin}$ = 62.9 min (Schwarzenberg-Czerny,\\ 
& Udalski \& Monier 1992).\\
DY Pup	&Observations too sparse to check for magnetic signature.\\
V1500 Cyg	&Polar.\\
V909 Sgr	&Not well observed.\\
RR Cha	&Intermediate polar with $P_{spin}$ = 32.50 min (Woudt \& Warner 2002).\\
RR Pic	&At one time thought to be an intermediate polar but not confirmed\\ 
& by later observations (Haefner \& Schoembs 1985).\\
V500 Aql	&Well observed but no magnetic signature.\\
\hline
\end{tabular}
\caption{Properties of Novae with Orbital Periods near 3.3 hours}
\label{tab3}
\end{table}

Although partly indirect, this evidence for several magnetic systems 
within the 3.2 h cluster is strong, though most are not desynchronised 
polars.
     
A comparison between the $P_{orb}$ distributions for polars and novae is 
shown in Figure~\ref{fig2}. The clustering near 3.3 h is seen in both subclasses, but 
is much tighter and more pronounced in the polars. Adding the few 
intermediate polars to the polars does not change the distribution 
noticeably (especially as the former also have a weak preference for periods 
around 3.3 h, 5 of about 15 with $P_{orb}$ < 5 h being close to that value). 
There are selection effects, depending on period, that distort these period 
histograms (Diaz \& Bruch 1997), but for comparisons between the subclasses 
these are probably not important.
    
Direct and indirect evidence for magnetic primaries also exists outside 
of the two period spikes. DQ Her and V533 Her are classic DQ Her stars. 
Confusion between the effects of asynchronism and those of superhumps leaves 
the statuses of V2214 Oph (Baptista et al. 1993) and GQ Mus (Diaz \& Steiner 
1989) uncertain. On the other hand, a high mass transfer disc with the 
centre missing is a characteristic of an intermediate polar, and such 
evidence exists for V Per (Shafter \& Abbot 1989). 
     
A possible prehistoric magnetic nova is the polar RX\,J1313.2-3259 
(G\"ansicke et al 2000), which, with the relatively long orbital period of 
4.19 h, would be expected to have a moderately high accretion rate and, from 
comparison with other CVs, a primary accretion-heated to about 30\,000 K. In 
fact, RXJ1313 has a measured temperature of 15\,000 K, which is compatible 
with heating at the estimated accretion rate of $\sim 10^{-10}$ M$_{\odot}$ y$^{-1}$. To cool 
and reach equilibrium at 15\,000 K after a nova eruption takes $\sim 10^4$ y, so 
G\"ansicke et al. suggest that the low accretion rate and temperature may be 
the result of an extended hibernation state that has lasted since a nova 
eruption $\sim 10^4$ y ago.

\begin{theacknowledgments}
Steve Potter kindly communicated orbital periods for some recently 
recognised polars. Patrick Woudt assisted with preparation of the paper and 
with helpful conversations. The author's research is funded by the 
University of Cape Town.
\end{theacknowledgments}

\bibliographystyle{aipprocl} 

\end{document}